\documentclass{article}
\usepackage{spconf,amsmath,graphicx,hyperref}

\usepackage{cite}
\usepackage{amssymb,amsfonts}
\usepackage{algorithmic}
\usepackage{textcomp}
\usepackage{xcolor}
\usepackage{booktabs, multicol, multirow, threeparttable}
\renewcommand{\paragraph}[1]{\vspace{1em}\noindent\textbf{#1}~~}

\usepackage{tipa}
\usepackage{makecell}

\definecolor{Highlight}{rgb}{0.12,0.49,0.85}
\usepackage{caption}
\usepackage{marvosym}

\newcolumntype{?}{!{\vrule width 0.8pt}}

\usepackage[capitalize]{cleveref}
\crefname{section}{Sec.}{Secs.}
\Crefname{section}{Section}{Sections}
\Crefname{table}{Table}{Tables}
\crefname{table}{Tab.}{Tabs.}

\definecolor{bl}{rgb}{0.25, 0.5, 0.9}

\usepackage{xspace}
\newcommand{\NAME}{S$^{3}$G\xspace}

\title{\NAME: Stock State Space Graph for Enhanced Stock Trend Prediction}
\name{Yao Lu$^{1}$ \quad Kaiyi Hu$^{2}$ \quad Luyan Zhang$^{3,*}$  \thanks{Corresponding author: zhang.luya@northeastern.edu}}

\address{$^{1}$University of Twente \qquad $^{2}$Georgia Institute of Technology \qquad $^{3}$Independent Researcher}

\begin{document}
%
\maketitle
\begin{abstract}
Stock trend prediction has attracted considerable attention for its potential to generate tangible investment returns. With the advent of deep learning in quantitative finance, researchers have increasingly recognized the importance of synergies between stocks, such as sector membership or upstream-downstream relationships, in accurately capturing market dynamics. However, previous work often relies on static industry graphs or constructs graphs at each time step via similarity measures, overlooking the fluid evolution of stock relationships. We observe that as companies interact competitively and cooperatively, their interdependencies change in a fine-grained, time-varying manner that cannot be fully captured by coarse, static connections or simple similarity-based snapshots. To address these challenges, we introduce the Stock State Space Graph (\NAME) framework for enhanced stock trend prediction. First, we apply wavelet transforms to denoise the inherently low signal-to-noise financial series and extract salient patterns. After that, we construct data-dependent graphs at each time point and employ state space models to characterize the evolutionary dynamics of these graphs. Finally, we perform a graph aggregation operation to obtain the predicted return. Extensive experiments on historical CSI 500 data demonstrate the state-of-the-art performance of \NAME, with superior annualized returns and Sharpe ratios compared to other baselines.
\end{abstract}

\section{Introduction}
\label{sec:intro}

Stock trend prediction, which leverages the historical data of all constituent stocks within a given universe to predict cross-sectional returns, aims to construct a portfolio capable of outperforming the market over time~\cite{fintsb,dai2024periodicity,timebridge}. This task is notoriously challenging due to the extremely low signal‑to‑noise ratio inherent in stock returns and the pronounced non‑stationarity of financial time series. Noise arising from myriad exogenous shocks, ranging from macroeconomic announcements to company-specific news, often obscures the true directional tendencies of price movements, while structural regime shifts continually alter the statistical properties of returns over time.

Most previous work has increasingly recognized that modeling stocks in isolation neglects the rich interdependencies between stocks~\cite{MDGNN}. Intuitively, shocks to one sector can trigger correlated ripples in adjacent industries, e.g., a sudden disruption in the supply of semiconductors will affect hardware manufacturers. Broadly speaking, methods for capturing such synergy effects fall into two classes. Some work constructs a static relationship graph using domain priors~\cite{hist}, predefined rules, or sector classifications. However, this fixed topology fails to reflect the evolving nature of inter‑stock linkages, e.g., competitors and collaborators within the same industry may exhibit opposite price movements on a given day. Another class infers dynamic correlations either via similarity metrics~\cite{tcgpn} or data‑driven learning-based methods~\cite{dai2024ddn,liu2024wftnet}, typically building a heterogeneous graph at each time slice. While this adaptability suits markets where the set of active stocks and their interactions change over time, slice‑by‑slice graphs struggle to capture delayed dependencies across multiple steps. For instance, upstream and downstream firms along a supply chain may react to the same macro factor with different lags, a pattern that cannot be easily captured in per-slice constructions.

To overcome these barriers, we present a novel paradigm that explicitly models the evolution of inter-stock relationships over time, rather than relying on static graph or slice similarity. By formulating a state space transition framework over a temporal graph, we learn how the connectivity at each step is shaped by the full trajectory of past interactions. This allows us to infer the precise moments when edges weaken or strengthen, capturing time-specific activations and decays to more accurately anticipate the topological structure at future horizons. Unlike methods that either freeze relationships or recalculate them purely from the latest observations, our approach tracks the latent influence of historical patterns on current couplings, yielding far greater flexibility in adapting to the dynamic nature of the financial market.

Technically, we introduce \NAME, which consists of two core components, a Wavelet Denoising Net (WDN) and a State Space Graph Learning (SSGL) module. First, to mitigate the effects of non-stationary noise, we apply discrete wavelet decomposition to isolate the principal fluctuation modes within each lookback window. We then partition the denoised series into consecutive time steps and construct a time-specific graph at each interval via Gaussian kernel graph construction on feature embeddings. Next, our state space graph transition equations ingest these sequential graphs to learn a parametric rule for how edge weights evolve, producing a predicted adjacency for the next time step. Finally, a graph neural network aggregates over this inferred structure to extract synergistic relationship factors, which drive cross-sectional return forecasts. Extensive experiments on the Chinese A-share market demonstrate that our \NAME consistently outperforms existing baselines across multiple evaluation metrics.

To summarize, our main contributions are as follows.
(1) We delve into existing approaches to capturing inter-stock synergistic effects and introduce a novel formulation of state space equations on dynamic stock graphs, enabling the unified modeling of both instantaneous and time-lagged dependencies.
(2) We propose \NAME, which employs wavelet transforms to denoise historical series and isolate core fluctuation patterns, then dynamically constructs time-specific relationship graphs to pinpoint time-varying edges.
(3) All components are seamlessly integrated and jointly to predict expected stock returns. Extensive evaluations on live Chinese A-share markets confirm that our \NAME achieves state-of-the-art performance for trend prediction.

\begin{figure*}[t]
    \centering
    \includegraphics[width=0.9\textwidth]{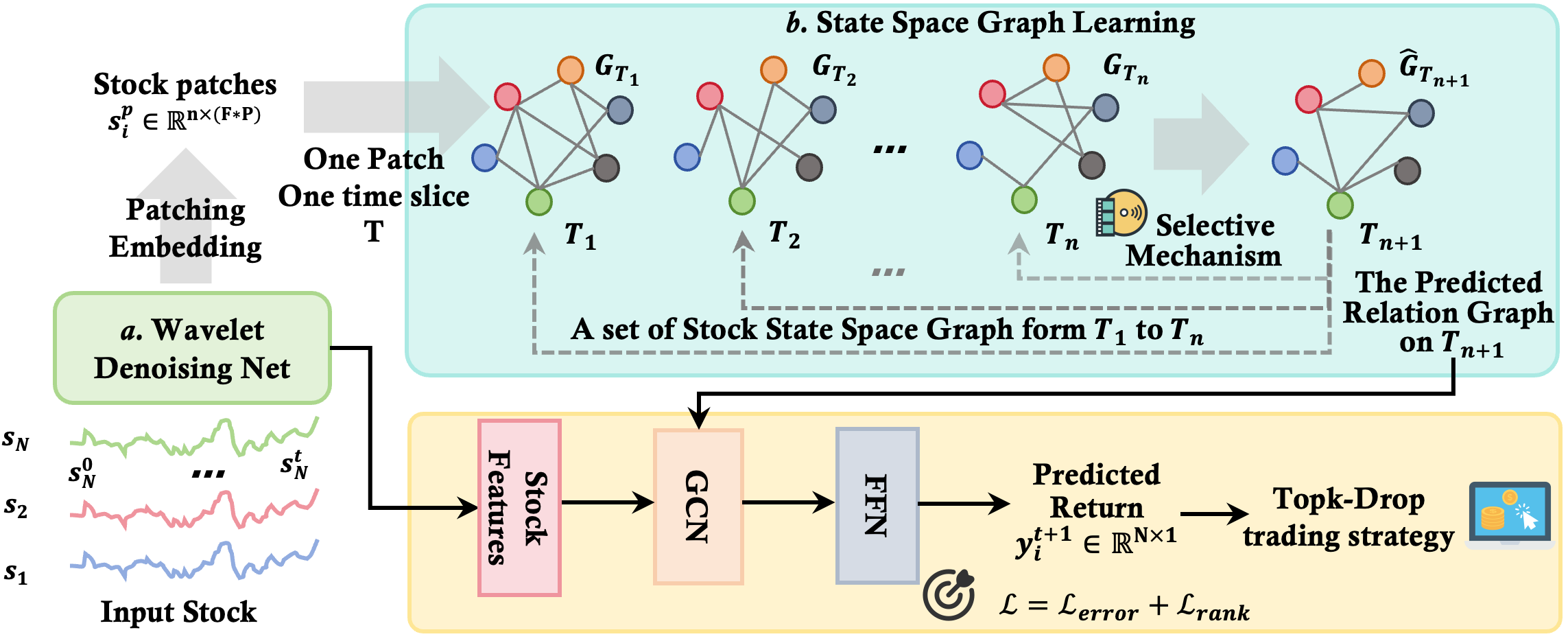}
    \caption{
    Overall structure of \NAME with (a) Wavelet Denoising Net and (b) State Space Graph Learning module.
    }
    \label{fig:pipeline}
\end{figure*}

\section{Background and Related Work}

\subsection{Graph Learning for Stock Trend Prediction}


Graph Neural Networks (GNNs) have gained attention~\cite{timefilter} in stock trend prediction for their ability to capture inter-stock dependencies. Conventional approaches such as GCN, GAT, and GraphSAGE rely on static graphs built from sector classifications, correlations, or fundamentals, while real-world relations are dynamic and time-varying. Temporal graph methods address this by incorporating evolving structures, e.g., TCGPN~\cite{tcgpn} uses industry affiliation and Euclidean distance, and CI-STHPAN~\cite{cisthpan} constructs hypergraphs via DTW. Yet, most still neglect history-dependent coupling dynamics. Our state space graph framework adaptively models both immediate and delayed interactions, producing fine-grained, time-specific structures that markedly improve return forecasts.

\subsection{State Space Models for Stock Trend Prediction}


State Space Models (SSMs) have recently shown strong performance in time series forecasting. In stock prediction, they are mainly used to model temporal dependencies within individual price histories, e.g. FinMamba~\cite{finmamba}. Unlike these single-series approaches, our \NAME extends state space reasoning to the graph domain, capturing the evolving inter-stock relationships and uncovering both immediate and lagged market influences.

\section{Method}

\subsection{Problem Statement}

The task of stock trend prediction can be framed as follows:
We have a collection of stock price sequences denoted as $\mathcal{S}=\{s_1, s_2, ..., s_N\}\in\mathbb{R}^{N\times L\times F}$, where each sequence $s_i$ corresponds to one stock, $N$ is the total number of equities, $L$ is the lookback window length and $F$ is the feature dimension.
For stock $s_i$ on trading day $t$, its feature vector is $s_i^t\in\mathbb{R}^F$, with the closing price $p_i^t$ among those features. We define the one-day return as $r_i^t = \frac{p_i^t-p_i^{t-1}}{p_i^{t-1}}$.
On any given day $t$, there exists an ideal score ordering $Y^t=\{y_1^t\geq y_2^t\geq...\geq y_N^t\}$.
such that for any two stocks $s_i, s_j\in\mathcal{S}$, $r_i^t\geq r_j^t$, implies $y_i^t\geq y_j^t$.
This ranking $Y_t$ represents the predicted hierarchy of expected returns, where higher‑scored stocks are forecasted to yield greater profits on day $t$.

\subsection{Wavelet Denoising Net}

To distill meaningful patterns from the noisy fluctuations inherent in raw stock sequences, we employ a trainable wavelet-based denoising network. Given an input sequence $s_i^t$, for stock $i$ over the lookback window on trading day $t$, the module first applies a one-dimensional convolution with filters of kernel size of $2F$ to decompose the signal into low frequency $L$ and high frequency $H$ components:
\begin{equation}
    [H,L]=\text{Conv1D}(s_i^t)\in\mathbb{R}^{L\times 2F}.
\end{equation}

We then perform element-wise soft thresholding with a learnable threshold $\gamma>0$ on the high-frequency branch to suppress residual noise:
\begin{equation}
    H'=\text{sign}(H)\text{max}(H-\gamma,0).
\end{equation}

The threshold $\gamma$ is learned during training, allowing the network to adaptively decide how much detail to prune. Finally, the denoised representation is reconstructed by reassembling the low- and filtered high-frequency components into embedding $Q$:
\begin{equation}
    Q=L+H'\in\mathbb{R}^{L\times F}.
\end{equation}

Afterwards, we construct the state space graph $\mathcal{G}$. We first segment $Q$ into $n$ patches along the time axis, each patch of length $P=[\frac{L}{n}]$. Then each patch is mapped to an embedded patch token $S^p\in\mathbb{R}^{N\times n\times D}$.
\begin{equation}
    s^p = \text{Patching}(s)\in\mathbb{R}^{N\times n\times(F*P)},\ \ 
    S^p = \text{Embedding}(s^p).
\end{equation}
The $\text{Embedding}(\cdot)$ operation transforms each patch from its original length $P$ to a hidden dimension $D$ through a trainable linear layer. In addition, we flatten the last two dimensions to obtain a set of temporal patches, where each patch contains a time slice information. There are all $n$ time slices, and we use $T_i,..,T_n$ to denote them.

Next, we construct a set of state space graphs $\mathcal{G}_{T_i},\forall i=\{1,2,...,n\}$ for each temporal patch $S^p$ by applying a Gaussian kernel similarity and get the adjacency matrix $A$ and $\mathcal{G}_{T_i}=\{S,A_{T_i}\}$ where $S$ is the node set with all the stocks and $A_{T_i}$ is the edge set on $T_i$.
\begin{equation}
    A_{T_i} = \text{Gaussian-kernel Sim}(S_{T_i}^p)\in\mathbb{R}^{N\times N}.
\end{equation}

\renewcommand{\arraystretch}{0.62}
\begin{table*}[!ht]
\setlength{\tabcolsep}{10pt}
\centering
\caption{Performance evaluation for stock trend forecasting in CSI 500. The best results are highlighted in \textbf{bold}.}
\resizebox{\textwidth}{!}
{
\begin{tabular}{c|ccccccccc}
\toprule

      

\multicolumn{1}{c}{} & \multicolumn{4}{c|}{\textbf{Ranking Metrics}} & \multicolumn{5}{c}{\textbf{Portfolio-Based Metrics}} \\

\multicolumn{1}{c}{\multirow{-2}*{\textbf{Methods}}} & \makebox[0.9cm]{IC$\uparrow$} & \makebox[0.9cm]{ICIR$\uparrow$} & \makebox[0.9cm]{RankIC$\uparrow$} & \multicolumn{1}{c|}{\makebox[1.2cm]{RankICIR$\uparrow$}} & \makebox[0.9cm]{ARR$\uparrow$} & \makebox[0.9cm]{AVol$\downarrow$} & \makebox[0.9cm]{MDD$\downarrow$} & \makebox[0.9cm]{ASR$\uparrow$} & \makebox[0.9cm]{IR$\uparrow$} \\
\midrule
XGBoost~\cite{xgboost} & 0.033 & 0.486 & 0.066 & 0.683 & 0.151 & 0.168 & -0.143 & 0.896 & 0.015 \\
DoubleEnsemble~\cite{DoubleEnsemble} & 0.036 & 0.434 & 0.027 & 0.745 & 0.134 & 0.171 & -0.137 & 0.781 & 0.031 \\
\midrule
LSTM~\cite{lstm} & 0.034 & 0.371 & 0.046 & 0.574 & 0.124 & 0.209 & -0.188 & 0.593 & 0.024 \\
ALSTM~\cite{alstm} & 0.029 & 0.330 & 0.031 & 0.471 & 0.119 & 0.192 & -0.203 & 0.613 & 0.029 \\
GRU~\cite{gru} & 0.027 & 0.411 & 0.054 & 0.812 & 0.223 & 0.163 & -0.092 & 1.265 & 0.017 \\
GAT-LSTM~\cite{gat} & 0.030 & 0.348 & 0.033 & 0.522 & 0.157 & 0.139 & -0.092 & 1.127 & 0.030 \\
Transformer~\cite{attention} & -0.007 & -0.056 & -0.014 & -0.097 & 0.127 & 0.265 & -0.272 & 0.480 & -0.075 \\
Mamba~\cite{mamba} & 0.026 & 0.295 & 0.051 & 0.544 & 0.126 & 0.228 & -0.157 & 0.552 & 0.025 \\
\midrule
PatchTST~\cite{patchtst} & -0.043 & -0.433 & 0.077 & 0.673 & 0.102 & 0.164 & \textbf{-0.084} & 0.621 & -0.043 \\
Crossformer~\cite{crossformer} & -0.096 & -0.524 & -0.234 & -0.109 & -0.092 & 0.266 & -0.253 & -0.347 & -0.096 \\
SegRNN~\cite{segrnn} & 0.011 & 0.145 & 0.013 & 0.160 & 0.108 & 0.165 & -0.091 & 0.655 & 0.011 \\
AMD~\cite{amd} & 0.023 & 0.247 & 0.056 & 0.565 & 0.146 & 0.155 & -0.116 & 0.941 & 0.033 \\
MambaStock~\cite{mambastock} & 0.028 & 0.298 & 0.044 & 0.516 & 0.123 & 0.201 & -0.139 & 0.612 & 0.028 \\
\midrule
DDPM~\cite{ddpm} & 0.026 & 0.451 & 0.023 & 0.413 & 0.198 & 0.084 & -0.147 & 2.353 & 0.026 \\
FactorVAE~\cite{FactorVAE} & -0.004 & -0.061 & -0.004 & -0.077 & 0.093 & \textbf{0.081} & -0.220 & 1.146 & -0.004 \\
\midrule
\NAME (Ours) & \textbf{0.041} & \textbf{0.531} & \textbf{0.074} & \textbf{0.892} & \textbf{0.247} & 0.185 & -0.136 & \textbf{1.336} & \textbf{0.040} \\
\midrule
\multicolumn{10}{c}{\textbf{Component ablation of \NAME in CSI 500.}} \\
\midrule
w/o WDN & 0.019 & 0.297 & 0.034 & 0.418 & 0.163 & 0.208 & -0.152 & 0.784 & 0.013 \\
w/o SSGL & -0.077 & -0.093 & -0.104 & -0.111 & 0.105 & 0.175 & -0.120 & 0.601 & -0.008 \\
\bottomrule
\end{tabular}
}
\label{tab:mainexp}
\end{table*}
\renewcommand{\arraystretch}{1.0}

\subsection{State Space Graph Learning}

Next, we model the temporal evolution of these per-slice graphs via a state space formulation. Let $A_{T_i}\in\mathbb{R}^{N\times N}$ be the adjacency at slice $T_i$, and introduce a latent state vector $h_{T_i}\in\mathbb{R}^{N\times N}$ for each slices. We posit the following transition and emission equations:
\begin{equation}
\begin{aligned}
    h_{T_i} &= \bar{A}_{T_i} h_{T_{i-1}}+\bar{B}_{T_i} A_{T_i},\\
    \hat{A}_{T_{i+1}} &= C_{T_i} h_{T_i},
\end{aligned}
\end{equation}
where $\bar{A}_{T_i}$, $\bar{B}_{T_i}$ and $C_{T_i}$ are learnable parameters that capture how past graph structures and current observations jointly influence the next graph. During training, these matrices adapt to fit the dynamic patterns of the financial market.

Once $\hat{\mathcal{G}}_{T_{i+1}}=\{S, \hat{A}_{T_{i+1}}\}$ is predicted for each slice, we employ a graph neural network to aggregate each stock’s neighbors under the inferred connectivity for stock $j$:
\begin{equation}
    z_{j}=\text{GNN}(\hat{\mathcal{G}}_{T_{i+1}}, S).
\end{equation}

This produces a set of relationship factors $z_{j}$ for each stock, where $S$ is the node features of all stocks. Finally, a feed‑forward network translates these factors into the next‑day return estimate:
\begin{equation}
    y_i^t = \text{FFN}(z_j).
\end{equation}

This end-to-end framework jointly learns denoising, graph evolution and return prediction, enabling fine-grained capture of both instantaneous and delayed inter-stock effects.

\subsection{Optimization Objective}

The training process employs a dual-objective optimization framework, simultaneously minimizing a composite loss function that integrates both Mean Squared Error (MSE) loss and ranking loss. 
These complementary objectives are balanced through a scaling coefficient $\eta$, which is set to $5.0$.
\begin{equation}
\begin{aligned}
    \mathcal{L}=&\frac{1}{L}\sum_{t=1}^{L}(\sum_{i=1}^{N}||y_i^t-r_i^t||^2+\\
    &\eta\sum_{i=1}^{N}\sum_{j=1}^{N}max(0,-(y_i^t-y_j^t)(r_i^t-r_j^t))).
\end{aligned}
\end{equation}

\section{Experiments}

\subsection{Experimental Settings}

\paragraph{\textbf{Datasets}}
We conduct experiments in the Chinese A-share stock markets. The datasets comprise historical day-level market information, including close, open, high, low, turnover, and volume.
We set the train set from 2015-01-01 to 2020-12-31, the validation set from 2021-01-01 to 2022-12-31, and the test set from 2023-01-01 to 2024-12-31.

\paragraph{\textbf{Parameter Settings and Environments}}
Our experiment is trained on a single NVIDIA 4090 GPU.
The training and validation sets are kept consistent for all models. The number of Mamba layers is $1$, the number of GNN layers is $1$, and the window size is $20$. The learning rate is $0.01$ and the weight of the ranking loss is set to $5.0$, while the weight of the MSE loss is set to $1.0$. We set $k$ to $10\%$ of the total number of stocks for the selected top $k$ ranked stocks in the experiment.

\paragraph{\textbf{Trading Strategy}}
Following~\cite{FactorVAE}, we adopt the Topk-Drop strategy to evaluate forecasting performance in terms of returns. On day $t$, an equal-weighted portfolio $\mathcal{P}^t$ of $m$ stocks is formed from the top-ranked predicted returns $Y$, with turnover constrained by $|\mathcal{P}^t\cap\mathcal{P}^{t-1}|\geq m-n$. In our setup, $m$ equals $10\%$ of all stocks and $n$ equals $3\%$, with a transaction cost of $0.1\%$ considered~\cite{fintsb}.

\paragraph{\textbf{Compared Baselines}}
We compare \NAME with other baselines from different categories as follows: 
(1) Classic investment strategies: Buying-Loser-Selling-Winner (BLSW)~\cite{blsw} and Cross-Sectional Mean reversion (CSM)~\cite{csm}; 
(2) Vanilla Deep Learning methods: LSTM~\cite{lstm}, ALSTM~\cite{alstm}, GRU~\cite{gru}, GAT~\cite{gat}, Transformer~\cite{attention}, Mamba \cite{mamba} and DDPM~\cite{ddpm};
(3) SOTA Deep Learning methods: PatchTST \cite{patchtst}, Crossformer~\cite{crossformer}, SegRNN \cite{segrnn}, AMD~\cite{amd}, MambaStock~\cite{crossformer}, and FactorVAE~\cite{FactorVAE}.

\paragraph{\textbf{Metrics}}
We employ wide-used two ranking metrics and five portfolio-based metrics to evaluate the overall performance of each method: Information Coefficient (IC), Information Coefficient Information Ratio (ICIR), RankIC, RankICIR, Annual Return Ratio (ARR), Annual Volatility (AVol), Maximum Draw Down (MDD), Annual Sharpe Ratio (ASR), and Information Ratio (IR). Lower absolute values of AVol and MDD, along with higher IC, ICIR, RankIC, RankICIR, ARR, ASR, and IR, indicate better performance.

\subsection{Main Results}

The overall performance is reported in \cref{tab:mainexp}. Our proposed \NAME outperforms all other methods on most metrics, confirming the benefit of explicitly modeling inter‑stock interactions. Notably, methods that ignore inter‑stock synergies lag significantly behind, underscoring the predictive value of capturing collaborative effects. Furthermore, while our model does not achieve the best AVol and MDD scores, reflecting the inherent trade-off between risk and return, it nevertheless secures the highest returns under comparable low-risk conditions, whereas the safest baselines sacrifice profitability for minimal volatility.

\subsection{Ablation Study}

The ablation study in the below of \cref{tab:mainexp} shows that omitting either the Wavelet Denoise Network (WDN) or the State Space Graph Learning (SSGL) module causes a dramatic drop in performance, highlighting the importance of both effective noise filtering and dynamic relationship modeling are for accurate stock trend prediction.

\section{Conclusion}

Existing stock trend prediction methods typically ignore the fine-grained evolution of inter-stock relations over time, focusing instead on static graphs or single-step similarities. 
To address this, we propose \NAME, which dynamically constructs time‑specific relationship graphs to identify and track evolving correlations.
Extensive experiments confirm the potential of our \NAME for capturing dynamic inter-stock behavior and improving forecast accuracy.

\newpage
\ninept
\bibliographystyle{IEEEbib}
\bibliography{strings,refs}

\end{document}